# Neutron-gamma discrimination by pulse analysis with superheated drop detector

Mala Das[1,3,*], S Seth[2,3], S Saha[1,3], S Bhattacharya[1,3], P Bhattacharjee[2,3]

[1] *Nuclear & Atomic Physics Division, Saha Institute of Nuclear Physics, Kolkata-700064, India*

[2] *Theory Division, Saha Institute of Nuclear Physics, Kolkata. 700064, India*

[3] *Centre for AstroParticle Physics, Saha Institute of Nuclear Physics, Kolkata, 700064, India*

*Corresponding author, E-mail: mala.das@saha.ac.in,

Tel : 91 33 2337 5346, Extn : 3304, Fax : 91 33 2337 4637

**Abstract** : Superheated drop detector (SDD) consisting of drops of superheated liquid of halocarbon is irradiated to neutrons and gamma-rays from $^{252}$Cf fission neutron source and $^{137}$Cs gamma source separately. The analysis of pulse height of the signals in the neutron and gamma-ray sensitive temperature provides strong information on the identification of neutron and gamma-ray induced events.

# 1. Introduction

Superheated drop detector (SDD) is known to detect neutrons, gamma-rays and other charged particles under different operating conditions [1-5]. It is also one of the promising detectors in search for WIMPs (Weakly Interacting Massive Particles), one of the favored candidates for cold dark matter [6-7]. The detector consists of drops of superheated liquid of low boiling point, suspended in a viscous gel or in a polymer medium. When radiation falls on the drops, if the energy deposition in the liquid exceeds the critical energy required for bubble nucleation, the bubbles are formed inside the drops. The critical energy depends on temperature and pressure of the liquid and on the types of sensitive liquid. The detector can be made insensitive to gamma-rays while sensitive to neutrons by varying the ambient temperature and pressure of the liquid, which essentially changes the threshold energy of detection. This property of the detector makes it an indispensable tool in neutron dose measurement in a strong background of gamma-rays. Another importance of the usefulness of neutron and gamma-ray discrimination lies in the WIMP search experiments using SDD. The strong possible candidates as the sources of background are neutrons, gamma-rays and alpha-particles. Therefore, development of efficient discrimination techniques is crucial for such experiments. PICASSO (Project In CAnada to Search for Supersymmetric Objects), one of the SDD based dark matter search experiments situated at SNOLab, Ontario, Canada, has recently set up a powerful discrimination technique for neutron and alpha-particle induced events in SDD with the sensitive liquid $C_4F_{10}$ [8]. It is found that alpha-particle induced signals are more intense than those due to neutrons. Photon sensitization temperature of different halocarbon and hydrocarbon and electron energy deposition were investigated by D'Errico, and an empirical formulation was established which states that the gamma-ray sensitive temperature would be the midpoint between the boiling point and the critical temperature of the liquid [3]. It also shows that the photon may trigger bubble nucleation via secondary electrons at their peak ionization density [3].

Gamma-ray sensitivity of R12 ($CCl_2F_2$ ; b.p. -29.8°C) was studied earlier using $^{137}$Cs, $^{241}$Am and $^{60}$Co sources, and that of R114 ($C_2Cl_2F_4$ ; b.p. 3.7°C) using $^{241}$Am gamma source, by varying the temperature. It was found that the sensitivity started at about 38.5°C for R12 and about 45°C for R114 in presence of $^{241}$Am, with the maximum rate at about 70°C [9-10]. The response rises slowly above the threshold. Another observation with R114 using $^{137}$Cs gamma source showed that the gamma-ray sensitization temperature was around 80°C [3]. Gamma-ray sensitivity for the detectors with sensitive liquids $CCl_2F_2$, $C_3F_8$, $C_4F_{10}$ and $C_2ClF_5$ were also studied by exposing to $^{57}$Co, $^{22}$Na, $^{137}$Cs and $^{60}$Co gamma sources. It was observed that the detectors became sensitive at around 34°C and reached an efficiency plateau between 40°C and 45°C [11].

In the above studies of gamma-ray sensitivities of various active liquids, the experiments were done by measuring the count rate as a function of temperature. But the detailed analysis of the pulse for each signal due to neutron and gamma-ray events has remained unexplored. This observation is important especially for those experiments where the event rate is very small, such as less than 1 event/kg/day of exposure for WIMP search experiments.

In the present work, the pulse height and power measurement at neutron and gamma-ray sensitive temperatures was done both with and without $^{252}$Cf fission neutron source. Separate experiments were also done with $^{137}$Cs gamma (662 keV) source to confirm the gamma-ray induced events. From the pulse height distribution and the power spectrum, the neutron and gamma-ray induced events can be discriminated. The acoustic signal of bubble nucleation was converted to electrical signal using condenser microphone and the analysis was done by recording the traces of the signals in digital storage Cathode Ray Oscilloscope. The manuscript is organized by explaining the experimental details, results coming out from the observations and discussions.

## 2. Experimental

The response of SDD to $^{252}$Cf was measured using superheated drops of R114 ($C_2Cl_2F_4$ ; b.p. $3.7^oC$) as the sensitive liquid suspended in an aquasonic gel matrix and by varying the temperature of the detector in the range of $35^oC$ to $75^oC$. The volume of the detector was 10ml in 15ml borosilicate glass vial and it was placed inside a water bath by controlling the temperature of the water with a temperature controller (Metravi, DTC-200) of precision $\pm1^oC$. The number of bubbles formed per minute was measured using an active drop counting device in the temperature range of room temperature to $75^oC$ [12]. The acoustic sensor part of the active drop counting device described in [12] has been used in the present work. The acoustic pulse associated with the process of bubble nucleation inside a liquid drop is converted to electrical signal using the condenser microphone based acoustic sensor. It contains two parallel plates in which one is movable. When acoustic signals fall on the movable plate, the distance between the plates changes and as a result its capacitance also changes. This gives rise to the corresponding electrical signal. The amplitude of the electrical signal depends on the intensity of the acoustic signal falling on it. The output of condenser microphone contains many undulations within the signal. The signal has a fast rise at the beginning and then falling slowly as damped oscillations. The traces of the pulses at 55°C and 70°C were measured using digital storage oscilloscope (MSO, Agilent Technologies, MSO7032A) both with and without the presence of $^{252}$Cf source. The oscilloscope was operated in 'single' trace mode with 25-30 mV of trigger level. The trace of the signal, recorded by oscilloscope was stored as ASCII data file which was exported to either EXCEL® or ORIGIN® program for plotting and further analysis. From the data file, the maximum positive amplitude of the waveform is noted. In separate experiments, the detector was irradiated with $^{137}$Cs (32.5 mCi) gamma-rays at 55°C and at and above 70°C, and traces of the pulses were recorded. Spontaneous nucleation at three different temperatures, 35°C, 45°C and 55°C using superheated drops of R12 ($CCl_2F_2$, b.p. -29.8°C) was observed and the pulses were recorded.

## 3. Results and Discussions

The variation of the normalized number of bubbles nucleated per minute with temperature of the detector observed earlier [12] is shown in Figure 1, which clearly indicates the existence of two distinct steps. The first step at the lower temperature is responsible for nucleation due to neutrons and the second one near 70°C is due to gamma-rays from $^{252}$Cf. The critical energy required for bubble formation decreases with increase in temperature of the detector. The energy deposition by neutrons is through production of heavy recoils, like C, Cl, F in the present case, and that by the gamma-rays is through the electrons. Energy deposition by the electrons in a critical length required for bubble nucleation is much lower than that by heavy recoils. Therefore, at lower temperature only the heavy recoils having higher LET (linear energy transfer) in the liquid can trigger bubble formation. At sufficiently higher temperature and correspondingly lower critical energy, gamma-rays having lower LET can also trigger bubble formation, which was already reported for this liquid [10]. The details of obtaining the neutron energy spectrum of $^{252}$Cf from Figure 1 was explained elsewhere [12]. The LET of C, Cl, F in R114 as a function of heavy recoils energy, the heavy recoils being produced via elastic head - on collision with neutron of energy 6 MeV which is the maximum significant energy in $^{252}$Cf, is calculated using the SRIM code [13], and is shown in Figure 2. The critical LET required for bubble nucleation at 55$^o$C in R114 is 254.65 keV/ μm with the nucleation parameter, k=0.0264 [10], which is found to be below the calculated LET of heavy recoils. Therefore, R114 is expected to be sensitive to neutrons from $^{252}$Cf at 55$^o$C which is also demonstrated by the experimental results. The measurement of the pulse height (H) distribution at 55$^o$C and 70$^o$C for $^{252}$Cf is shown in Figure 3. The result for $^{137}$Cs gamma-ray irradiation at and above 70$^o$C is shown in Figure 4. In Figure 3 and Figure 4, the maximum amplitude of the trace of each signal is plotted. The typical waveform for neutron and gamma-ray induced events are shown in Figure 5 and Figure 6, respectively.

The measurement at 55$^o$C with $^{252}$Cf is done to observe the pulse height (H) distribution due to neutrons only. Figure 3 shows that the pulses due to neutrons are of higher amplitude mainly in the pulse bin of 420-520mV and the pulses due to gamma-rays are in the lower amplitude, mainly in the 20-70mV amplitude bin. The observed pulses due to spontaneous nucleation at 70$^o$C are in between the pulses due to neutrons and gamma-rays. Spontaneous nucleation at 55$^o$C is negligible for R114 sample used in the present experiment. It is known that the spontaneous nucleation occurs in such detector in absence of any known radioactive source, and the rate of spontaneous nucleation increases with increase in temperature until limit of superheat is reached [14]. As temperature increases, the thermal fluctuations in the droplet can cause bubble nucleation. Therefore, at 55 $^{o}$C which is much below the critical temperature ($T_c$ = 145.7 $^{o}$C) of R114, the spontaneous nucleation rate is very low but at 70 $^{o}$C, the rate becomes significant as it reaches closer to the critical temperature.

Figure 4 shows that the pulses due to $^{137}$Cs at 70$^{o}$C and 75$^{o}$C are mainly of lower amplitudes. For $^{137}$Cs no response was observed at 55$^{o}$C. This is due to the fact that the detector is insensitive to gamma-rays at 55$^{o}$C. It is also observed that count rate is higher at 75$^{o}$C than that at 70$^{o}$C. This is because the nucleation probability for gamma-rays increases with temperature. This observation also confirms that the gamma-ray induced events are of lower amplitude. In another experiment with R12 detector, shown in Figure 7, it is observed that the pulse amplitude due to spontaneous nucleation increases with increase in temperature.

The measurement of maximum amplitude of the pulse provides rough information on the energy deposited and released during the nucleation process. A more precise way of presenting the results is in terms of power. The power is proportional to the energy released during the bubble formation process. Therefore, the results are also displayed in Figure 8 in terms of power, P [15]. P is defined as $Log_{10}\left(\sum_{i}|V_i|^2\right)$ where $V_i$ is the pulse amplitude in volt of the digitized acoustic pulse (e.g. Fig.5) at the i$^{th}$ time bin, and the summation extends over the duration of the signal. In the figure, P is rescaled to P = 0 corresponding to the lowest value. In Figure 8, the events are displayed for neutron and gamma-ray induced cases which show that the neutron induced events are of higher P than the gamma-ray induced events. As explained earlier, neutrons produce heavy recoils which deposit their energy in the medium with higher LET. For gamma-rays, it is the electrons that deposit the energy in the medium almost at the end of their track [3], which is much smaller in a critical diameter than that by heavy recoils. The pulse height and the power spectrum are dependent on the energy deposited and released during bubble formation. Having higher LET values for heavy recoils, neutron induced events show higher P than the gamma-ray induced events.

These important observations with SDD clearly demonstrates for the first time, the identification of neutrons and gamma-rays by measuring the pulse height distribution in a mixed radiation field, such as in neutron dosimetry and also in WIMP search experiments using SDD.

## Acknowledgment

Mala Das is funded by Department of Science and Technology, DST, Govt of India, WOS Scheme-A. Authors are grateful to Prof. B. K. Chatterjee, Dept of Physics, Bose Institute for providing R114 liquid and $^{137}$Cs gamma source. One of the authors (M.D) thanks Prof. Viktor Zacek, University of Montreal, Canada, for valuable discussions and communications.

**Figure captions**

Figure 1. Experimentally observed variation of normalized number of bubbles per minute with temperature of the detector liquid.

Figure 2. Calculated value of LET $\left(\frac{dE}{dx}\right)$ of heavy recoils, C, Cl, and F in R114 using SRIM 2008 code along with the critical LET at 55°C.

Figure 3. Experimentally observed differential pulse height (maximum) distribution (variation of number of pulses) as a function of pulse height (H) in presence of $^{252}$Cf at 55°C and 70°C and also for spontaneous nucleation at 70°C.

Figure 4. Observed differential pulse height (maximum) distribution in presence of $^{137}$Cs at 70°C and at 75°C.

Figure 5. Typical waveform of the condenser microphone output for neutron induced nucleation events.

Figure 6. Typical waveform of the condenser microphone output for gamma-ray induced nucleation events.

Figure 7. Observed variation of differential pulse height (maximum) distributions for spontaneous nucleation at different temperatures for R12.

Figure 8. Power distribution of the pulses as recorded with $^{252}$Cf at 55°C, 70°C and with $^{137}$Cs at 70°C.

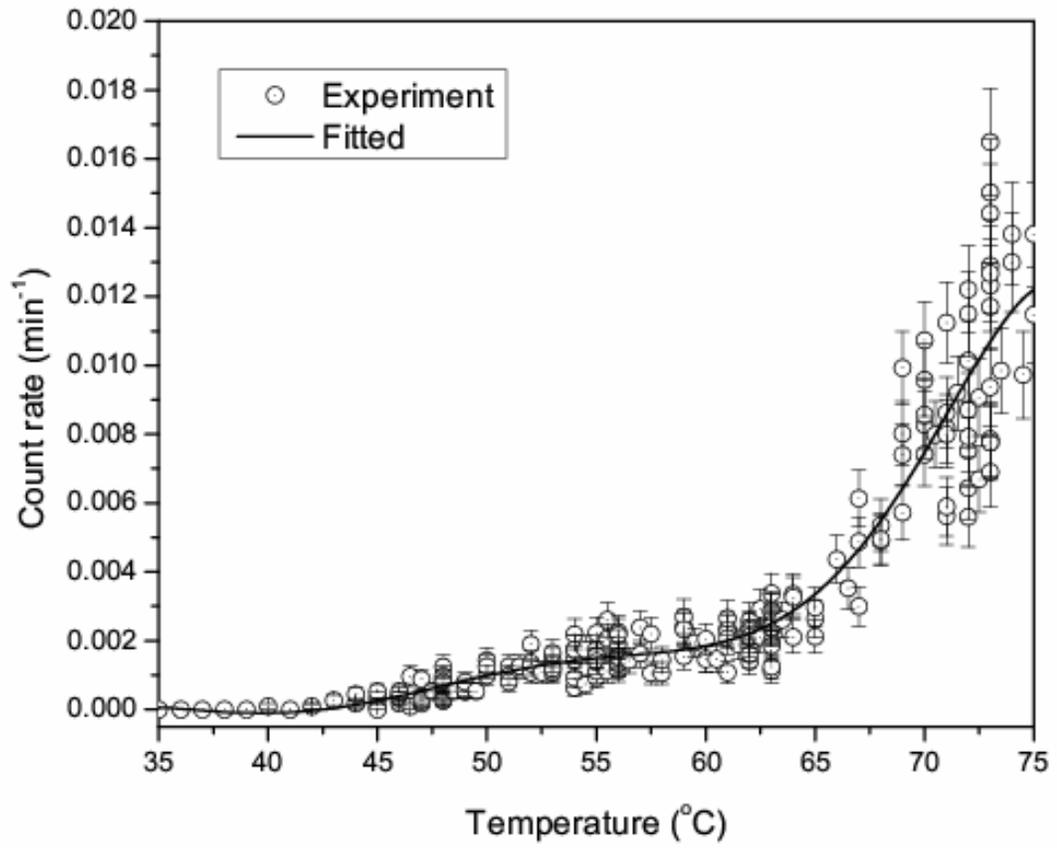

Figure.1

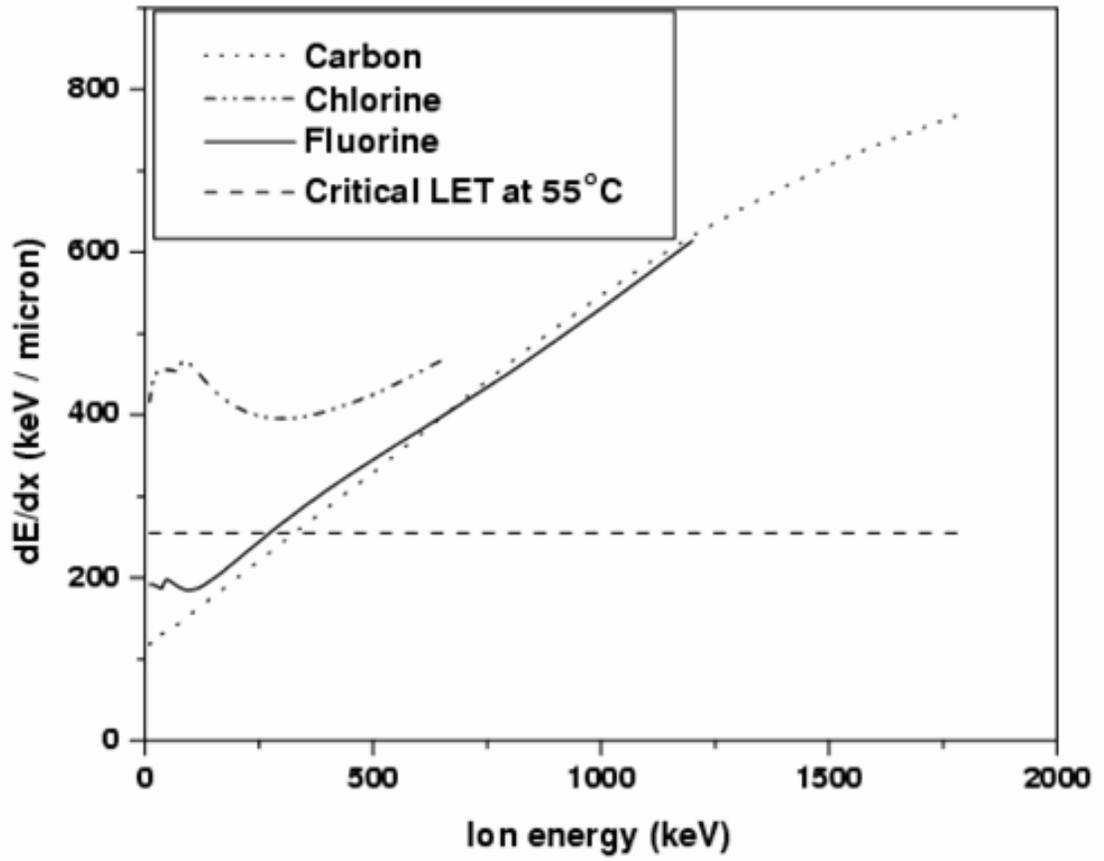

Figure.2

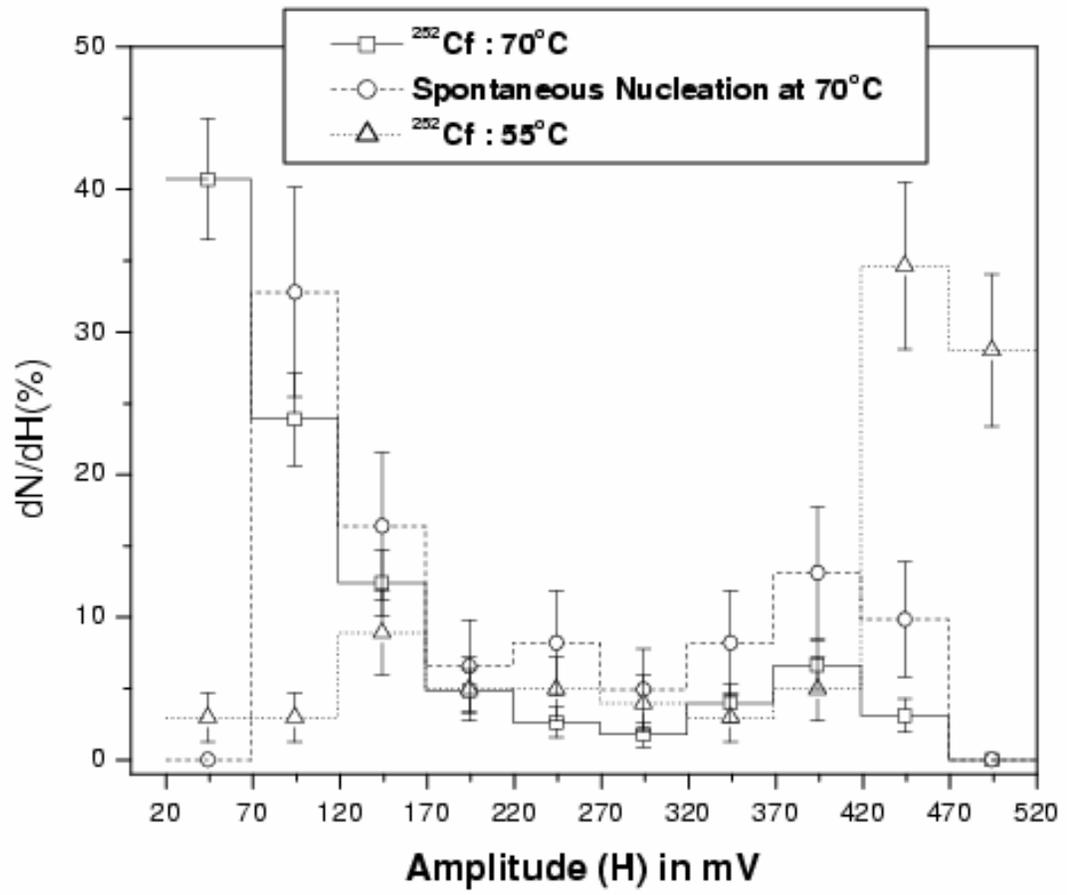

Figure.3

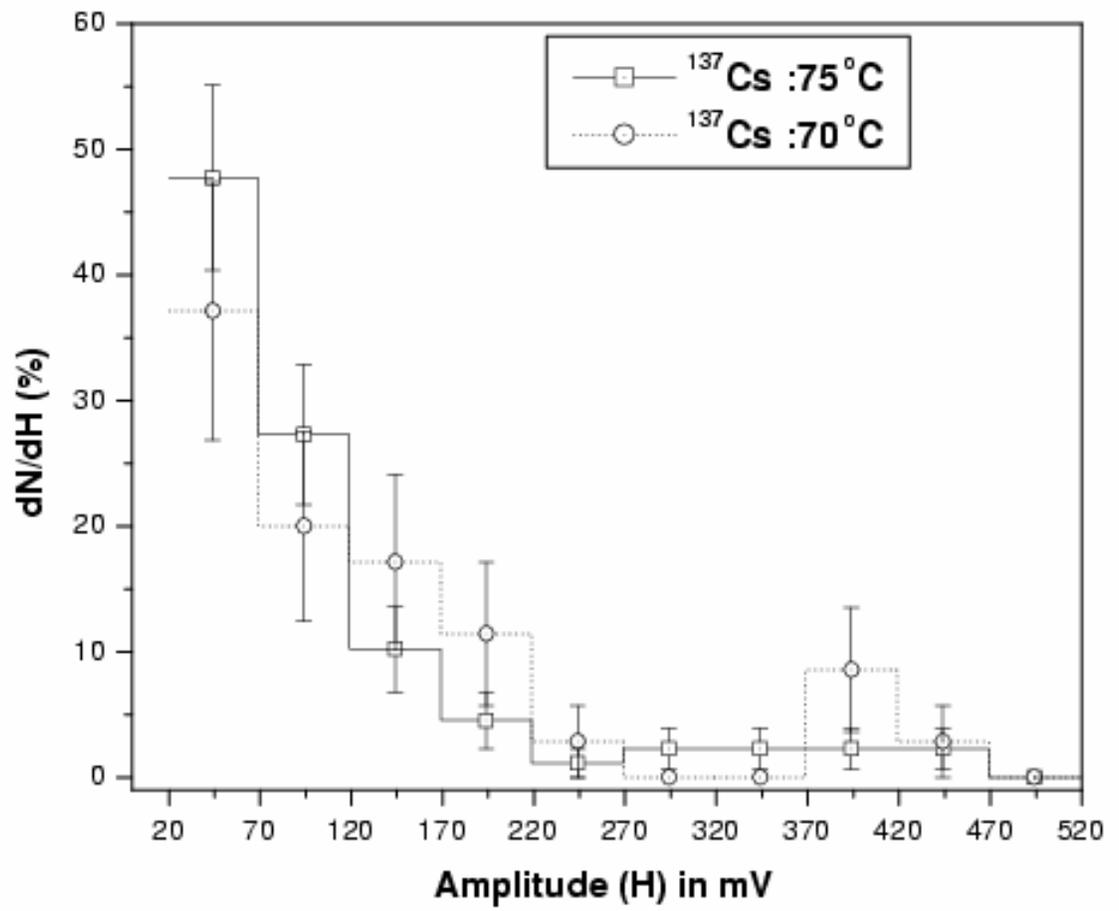

Figure.4

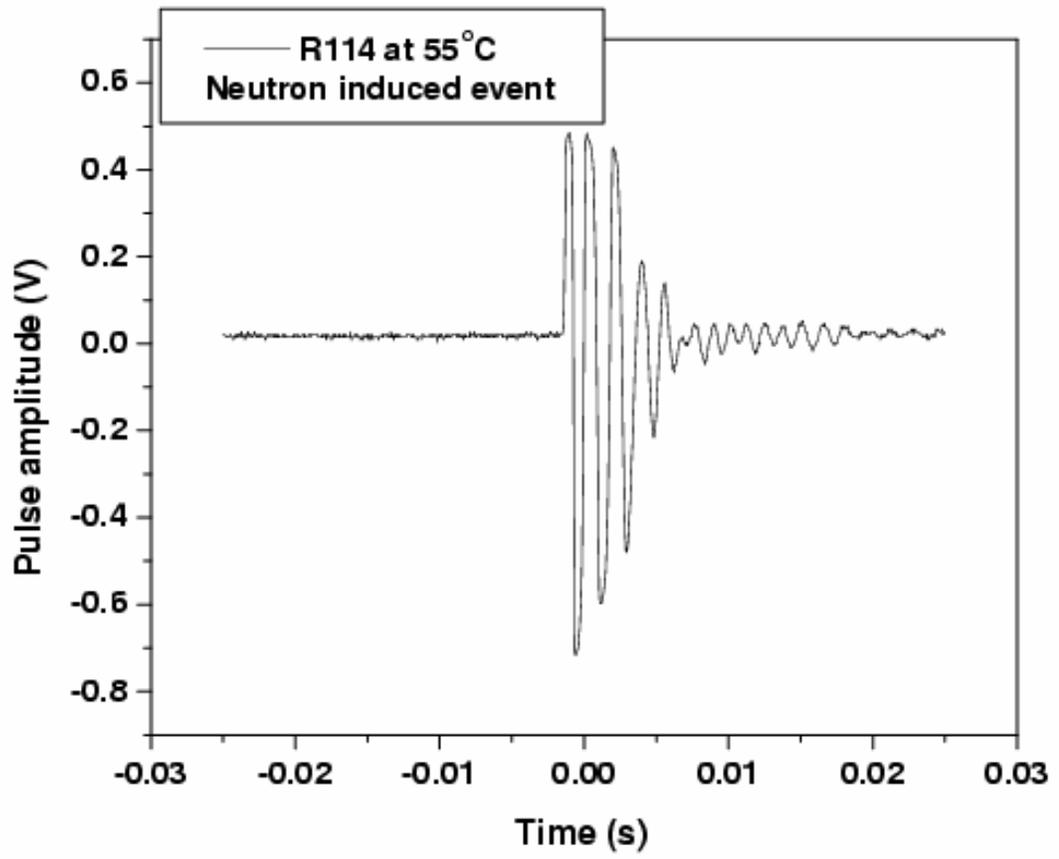

Figure.5

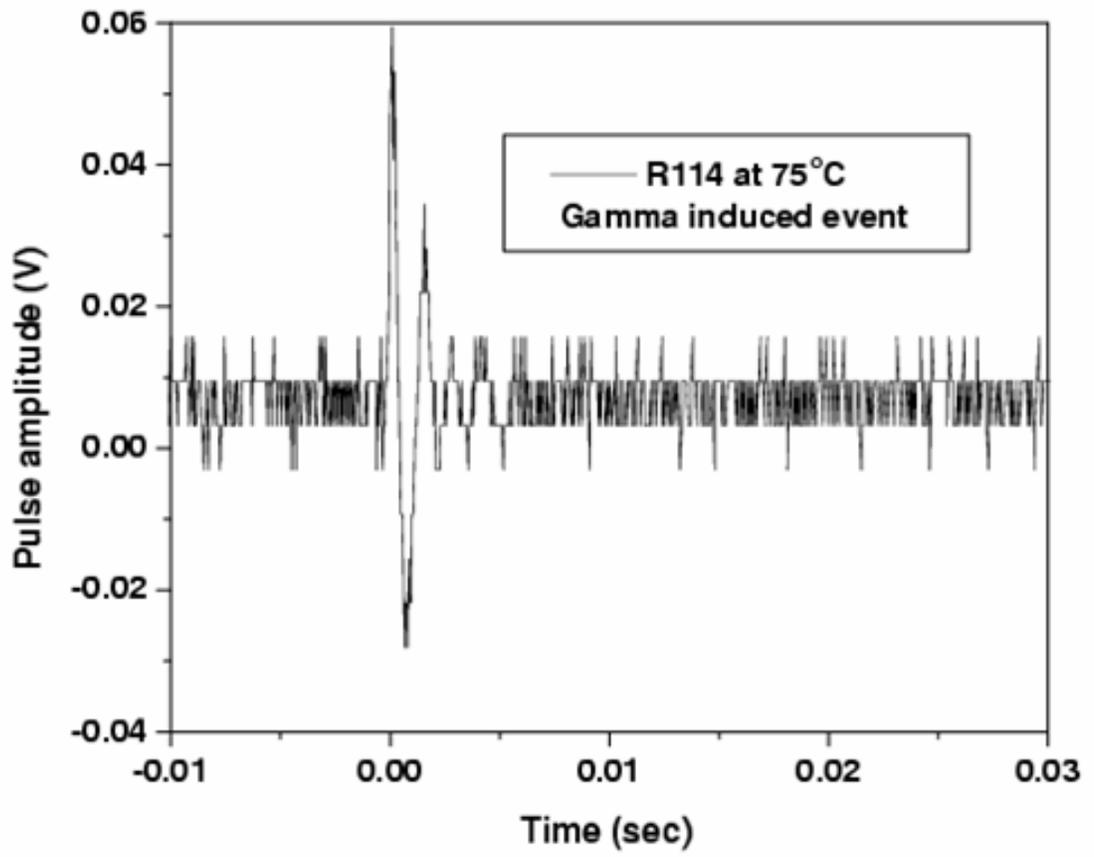

Figure.6

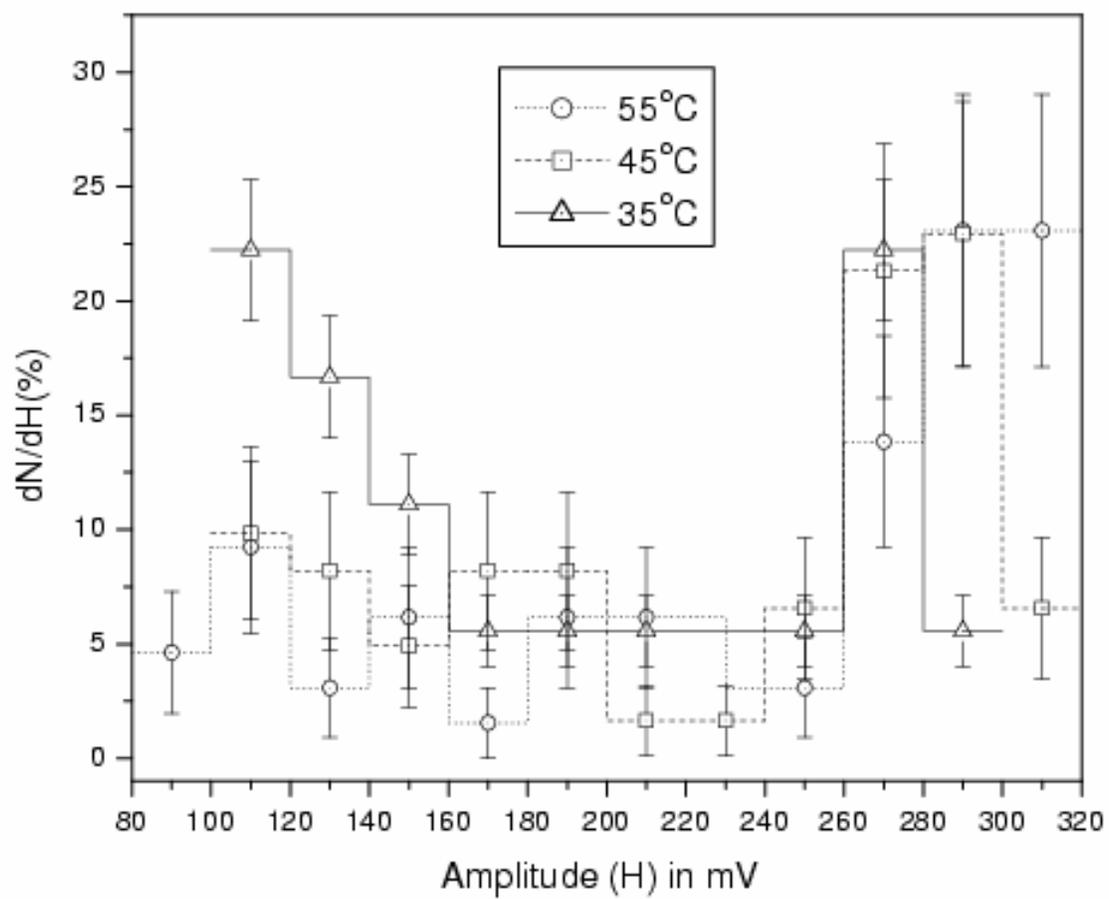

Figure.7

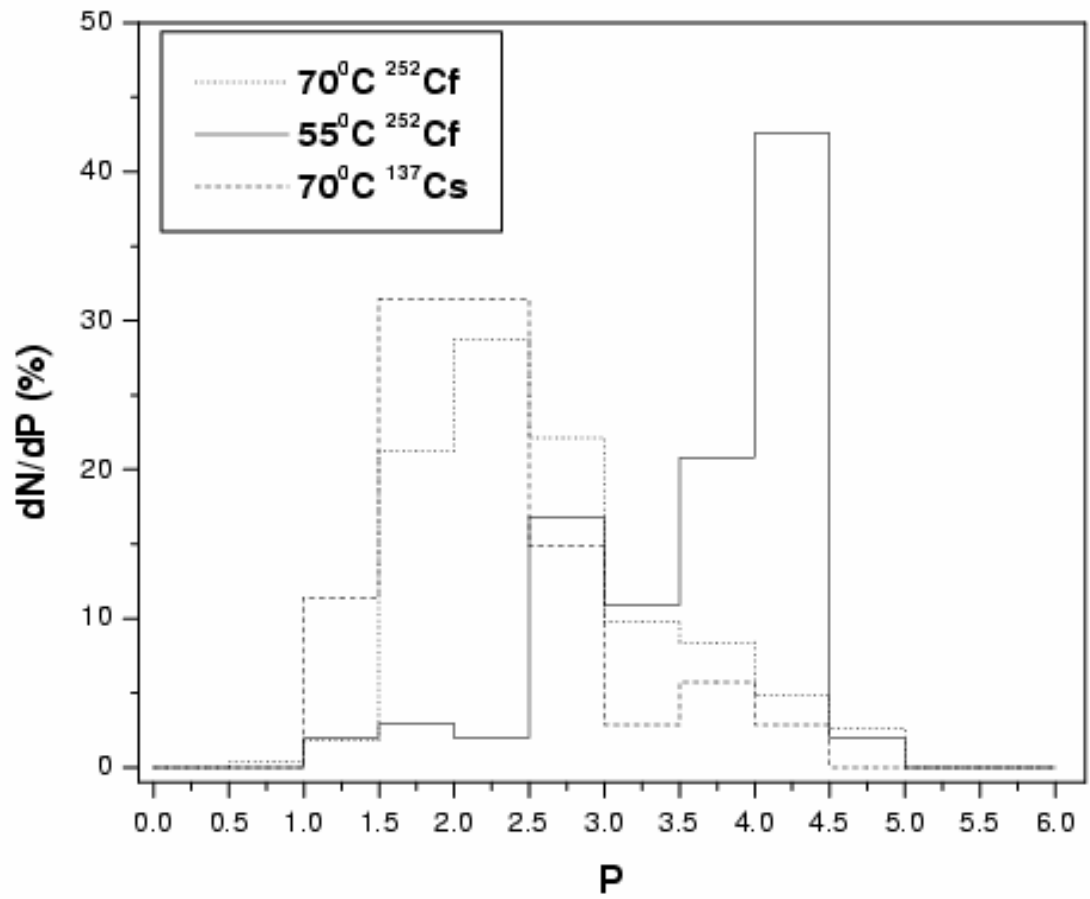

Figure.8